\documentstyle[aps,graphicx,epsf]{revtex}\tighten
\def \BE {\begin{equation}}
\def \EE {\end{equation}}

\def \eq#1 {\eqno{(#1)}}

\def \e#1 {{\rm e}^{#1}}

\def\vell {\bf{\ell}}
\def\vm {\bf{m}}
\def\vn {\bf{n}}

\def \tT{\bf{T}}

\def\bo {\cal{B}}

\begin{document}

\title{Classification of the Weyl Tensor in Higher Dimensions}  
\author{A. Coley\dag, R. Milson\dag, V. Pravda\ddag, and  A. Pravdov\' a\ddag} 
\address{\dag\ Department of Mathematics and Statistics,
Dalhousie University, Halifax, Nova Scotia}
\address{\ddag\ Mathematical Institute, 
Academy of Sciences, \v Zitn\' a 25, 115 67 Prague 1, Czech Republic\\
 {aac@mathstat.dal.ca},   {milson@mathstat.dal.ca},
  {pravda@math.cas.cz},  {pravdova@math.cas.cz}}

\maketitle

\begin{abstract}

We discuss the algebraic classification of the Weyl tensor in
higher dimensional Lorentzian manifolds. This is done by
characterizing algebraically special Weyl tensors by means of the
existence of aligned null vectors of various orders of alignment.
Further classification is obtained by specifying the alignment
type and utilizing the notion of reducibility. For a complete
classification it is then necessary to count aligned directions,
the dimension of the alignment variety, and the multiplicity of
principal directions. The present classification reduces to the classical
Petrov classification in four dimensions. Some applications are briefly discussed.

\end{abstract}

 \vskip .15in
 
\noindent 
[PACS: 04.20.J, 98.90]

\vskip .1in

The study of higher dimensional manifolds in gravity theory is
currently of great interest \cite{wesson,soliton,nrbh}. In particular, finding
spacetime models in higher dimensions in the context of string
theory, especially in ten and eleven dimensional  supergravity
theories \cite{examples} and branes embedded in higher dimensional bulk manifolds
\cite{brane}, is of fundamental importance. Therefore, a
mathematical classification of higher dimensional manifolds is of
particular importance. Recently, the algebraic structure of 
tensors in higher dimensional Lorentian manifolds, including the
Riemann and Weyl tensors, 
has been discussed \cite{MCPP}. Here we shall concentrate on the
Weyl tensor and present a higher dimensional algebraic classification
which is a generalization of the
Petrov classification in four dimensions \cite{petrov}.

We shall consider a null frame  $\vell=\vm_0,\;\vn=\vm_1,\;\vm_2,...\vm_i$
($\vell,\;\vn$ null with  $\ell^a \ell_a = n^a n_a = 0, \ell^a n_a = 1$,
$\vm^i$ real and spacelike with $m_i{}^a m_j{}_a = \delta_{ij}$; all other products vanish)
in an  $N$-dimensional Lorentz-signature space(time),
so that $g_{ab} = 2l_{(a}n_{b)} +  \delta_{jk} m^j_a m^k_b$.
Throughout, Roman indices $a,b,c, A,
B,C$ range from $0$ to $N-1$.  Lower case
indices indicate an arbitrary basis, while the upper-case ones
indicate a null frame. Space-like indices $i,j,k$ also indicate a
null-frame, but vary from $2$ to $N-1$ only.   We will use
Einstein's summation convention for both  of these types of
indices; however, note that for indices $i,j \dots$ there is no difference
between covariant and contravariant components and thus we will
not distinguish between subscripts and superscripts.

The frame is covariant relative to the group of linear
Lorentz transformations. 
A {\em null rotation} about $\vn$ is a
Lorentz transformation of the form
\begin{equation}
  \label{eq:nullrot}
    \hat{\vn}=  \vn,\quad
    \hat{\vm}_i=  \vm_i + z_i \vn,\quad
    \hat{\vell}= \textstyle \vell
    -z_i \vm^i-\frac{1}{2} \delta^{ij} z_i z_j\, \vn.
\end{equation}
A null rotation about $\vell$ has an analogous form.  A boost is a
transformation of the form
\begin{equation}
  \label{eq:boost}
    \hat{\vn}= \lambda^{-1}\vn,\quad
    \hat{\vm}_i=  \vm_i,\quad
    \hat{\vell}=  \lambda\, \vell, \quad \lambda \neq 0.
\end{equation}
A spin is a transformation of the form
\begin{equation}
  \label{eq:spin}
    \hat{\vn}= \vn, \quad
    \hat{\vm}_i=  X_i^j\,\vm_j,\quad
    \hat{\vell}=  \vell,
\end{equation}
where $X_i^j$ is an orthogonal matrix.

Let $T_{a_1... a_p}$ be a rank $p$ tensor.  For a fixed list of
indices $A_1,...,A_p$, we call the corresponding $T_{A_1... A_p}$
a null-frame scalar.  These scalars transform under a boost
(\ref{eq:boost}) according to
\begin{equation}
  \label{eq:boostxform}
  \hat{T}_{A_1... A_p}= \lambda^b\,
  T_{A_1... A_p},\quad b=b_{A_1}+...+b_{A_p},
\end{equation}
where $b_0=1,\quad b_i=0,\quad b_1=-1$. We call the
above $b$ the boost-weight of the scalar.
We define the {\em boost order} of the tensor $\tT$ to be
the boost weight of its leading term.

We introduce the notation \BE
T_{\{pqrs\}}\equiv\frac{1}{2}(T_{[ab][cd]}+
 T_{[cd][ab]}). \EE
We can decompose the Weyl tensor and sort the components of the
Weyl tensor by boost weight \cite{MCPP}:

\begin{eqnarray}
 C_{abcd} = {\overbrace{4C_{0i0j} n_{\{a}{m^i}_b n_c {m^j}_{d\}}
}}^2 + {\overbrace{8C_{010i} n_{\{a} \ell_b n_c {m^i}_{d\}}
 + 4C_{0ijk} n_{\{a} {m^i}_b {m^j}_c {m^k}_d\}}}^1 + && \nonumber \\
 \left\{\begin{array}{l}
 4C_{0101} n_{\{a} \ell_b n_c \ell_{d\}} + 4C_{01ij}
 n_{\{a} \ell_b {m^i}_c {m^j}_{d\}} +\label{eqnweyl} \\
 8C_{0i1j} n_{\{a} {m^i}_b \ell_c {m^j}_{d \}} +
 C_{ijkl} m^i_{\{a} {m^j}_b  {m^k}_c {m^l}_{d\}}\end{array}
 \right\}^0 + && \\
 {\overbrace{8C_{101i} \ell_{\{a} {n}_b \ell
 _c {m^i}_{d\}} + 4C_{1_{ijk}} \ell_{\{a} {m^i}_b {m^j}_c {m^k}_{d \}}}}^{-1}
 + {\overbrace{4C_{1i1j} \ell_{\{ a} {m^i}_b \ell_c {m^j}_{d\}}}}^{-2}.&& \nonumber
 \end{eqnarray}

The boost weights for the scalars of the Weyl curvature tensor is
given explicitly in (\ref{eqnweyl}) (this is summarized concisely in Table I in
\cite{MCPP}). The Weyl  tensor is generically of boost order
$2$. If all $C_{0i0j}$ vanish, but some $C_{010i}$, or $C_{0ijk}$
do not, then the boost order is $1$, etc. The Weyl scalars also
satisfy a number of additional relations, which follow from
curvature tensor symmetries and from the trace-free condition:
\begin{eqnarray}
C_{0i0}{}^i = 0, C_{010j} = C_{0ij}{}^i,  C_{0(ijk)} = 0, C_{0101}
= C_{0i1}{}^i,C_{i(jkl)} = 0,\nonumber\\
C_{0i1j}=-\frac{1}{2} C_{ikj}{}^k+\frac{1}{2}C_{01ij},  C_{011j} = -C_{1ij}{}^i,  C_{1(ijk)} = 0,  C_{1i1}{}^i =
0\label{cons}.
\end{eqnarray}
A null rotation about $\vell$ fixes the leading terms of a tensor,
while boosts and spins subject the leading terms to an invertible
transformation.  It follows that the boost order of a tensor is a
function of the null direction $\vell$ (only). We shall therefore
denote boost order by $\bo(\vell)$ \cite{MCPP}. We will {\em define}
a null vector $\vell$ to be {\em aligned} with the Weyl tensor 
whenever $\bo(\vell)\leq 1$ (and we shall refer to $\vell$ as a Weyl
aligned null direction (WAND)). We will call the integer
$1-\bo(\vell)\in \{0,1,2,3\}$ the order of alignment. The alignment 
equations are $\frac{1}{2}N(N-3)$ degree-4 polynomial 
equations in $(N-2)$ variables, which are in general overdetermined
and hence have no solutions for $N>4$ (i.e., only when these equations are degenerate
do we obtain algebraically special spacetimes). Grobner bases
can be efficiently used to  study this variety and determine any WANDs.

\noindent
{\bf Principal Classification:}
\newline
Following \cite{MCPP}, we will say that the {\bf principal type}
of a Lorentzian manifold is {\bf I, II, III, N} according to
whether there exists an aligned $\vell$ of alignment order $0,1,2,3$
(i.e. $\bo(\vell)=1,0,-1,-2$), respectively. If no aligned $\vell$
exists we will say that the manifold is of (general) type {\bf G}.
If the Weyl tensor vanishes, we will say that the manifold is of
type {\bf O}. The algebraically special types are summarized as
follows (using (\ref{cons})):
\begin{eqnarray}
&&  Type ~~{\bf I}: ~~ C_{0i0j}=0 \nonumber\\
&&  Type~~ {\bf II}: ~~C_{0i0j}=C_{0ijk}=0 \nonumber \\
&&  Type ~~{\bf III}:  ~~C_{0i0j}=C_{0ijk}=C_{ijkl} =C_{01ij}=0 \nonumber\\
&&  Type ~~{\bf N}:  ~~C_{0i0j}=C_{0ijk}=C_{ijkl} = C_{01ij}=C_{1ijk}=0
\end{eqnarray}

\noindent
{\bf Secondary Classification:}
\newline
(i) The existence of aligned null vectors of various orders of
alignment can be used to covariantly characterize algebraically
special Weyl tensors. Further categorization can be obtained by
specifying {\em alignment type} \cite{MCPP}, whereby we try to
normalize the form of the Weyl tensor by choosing both $\vell$ and
$\vn$ in order to set the maximum number of leading and trailing
null frame scalars to zero. Let $\vell$ be a WAND whose order of
alignment is as large as possible. We then define the principal (or primary)
alignment type of the tensor to be $b_{max} - b(\vell)$. Supposing
such a WAND $\vell$ exists, we then let $\vn$ be a null vector of
maximal alignment subject to $\ell_{a} n^{a}=1$. We define the
secondary alignment type of the tensor to be $b_{max}-b(\vn)$. The
alignment type of the Weyl tensor is then the pair consisting of
the principal and secondary alignment type. The possible alignment
types of a  higher dimensional Weyl tensor are categorized in
Table 4 of \cite{MCPP}. In general, for types ${\bf I}, {\bf II},
{\bf III}$ there does not exist a secondary aligned $\vn$ (in
contrast to the situation in four dimensions), whence the
alignment type consists solely of the principal alignment type.
Alignment types (1,1), (2,1) and (3,1) therefore form
algebraically special subclasses of types ${\bf I}, {\bf II}, {\bf
III}$ respectively (denoted types ${\bf I}_i, {\bf II}_i, {\bf
III}_i$). There is one final subclass possible, namely type (2,2)
which is a further specialization of type (2,1); we shall denote
this as type ${\bf II}_{ii}$ or simply as type ${\bf D}$.
Therefore, a type ${\bf D}$ Weyl tensor in canonical form has no
terms of boost weights $2,1,-1,-2$ (i.e., all terms are of boost
weight zero for type ${\bf D}$).

\noindent (ii) The above classification(s) can be utilized for arbitrary 
Lorentzian manifolds. In the case in which
the Weyl tensor is reducible, it is possible to obtain much more information
by decomposing the Weyl tensor and classifying
its irreducible parts. For example, we shall say that  $C_{abcd}$ is
{\em reducible} if there exists a null frame and a constant $M<N$
such that
\begin{equation}
\label{decomp}
 {^N}C_{abcd} = {^M}C_{abcd}+ {\tilde{C}}_{{\tilde
a}{\tilde b}{\tilde c}{\tilde d}} \end{equation}
where the only non-vanishing components of ${^N}C_{abcd}$ are
$${^M}C_{\scriptscriptstyle abcd} \neq 0,\quad 0\leq a,b,c,d\leq M-1;\quad
{\tilde{C}}_{\scriptscriptstyle {\tilde a}{\tilde b}{\tilde c}{\tilde d}} \neq 0,
  \quad M\leq {\tilde a},{\tilde b},{\tilde c},{\tilde d}\leq N-1.$$
That is, $C_{abcd}$ is (algebraically) reducible if and only if it is the sum of
two Weyl tensors. A reducible Weyl tensor is said to be (geometrically)
decomposable if and only if the components belonging to
$M_M$(resp. $\tilde{M}_{\tilde{M}}$) depend on only $x^c$ (resp.
$y^{\tilde c}$) \cite{KN}: ${^M}C_{abcd}$ is a Weyl tensor of an ({\em
irreducible}) Lorentzian spacetime of dimension $M$, 
${\tilde{C}}_{\scriptscriptstyle {\tilde a}{\tilde b}{\tilde
c}{\tilde d}}$ is a Weyl tensor of a Riemannian space of dimension
$N-M$, and we can write symbolically: $C_N = C_M \oplus
\tilde{C}_{\tilde{M}}$. We can also write a  decomposable Weyl
tensor in the suggestive block diagonal form: ${\bf block diag}\{
C_{abcd}(x^e) ,{\tilde C}_{{\tilde a}{\tilde b}{\tilde c}{\tilde
d}}(y^{\tilde e})\}$. This can be trivially generalized to the
case where $\tilde{M}_{\tilde{M}}$ is further reducible.
Writing out the Weyl tensor in terms of boost weights we obtain
(\ref{eqnweyl}), with indices $a,b,c,..$ running from $0 - (M-1)$,
and an additional term
\begin{equation}
(+) ~~~ \{\tilde{C}_{\tilde{i} \tilde{j} \tilde{k}\tilde{l}} 
m^{\tilde{i}}_{\{\tilde{a}}m^{\tilde{j}}\ \!_{\tilde{b}} 
m^{\tilde{k}}\ \!_{\tilde{c}} m^{\tilde{l}}\ \!_{\tilde{d}\}}    \}^0.
\end{equation}
This term, corresponding to the Riemannian part of the Weyl tensor
${\tilde C}$, is either identically zero (of type ${\bf O}$) or has terms of
boost weight zero only and hence is of
type ${\bf D}$.



An important auxiliary question is whether we can deduce anything
about the manifold structure if the Weyl tensor is reducible. For
example, if $M$ is a paracompact, Hausdorff, simply connected
smooth Lorentzian manifold,  in the case of vacuum it follows
that if the Weyl tensor is reducible then it is also decomposable,
whence there exists a real non-trivial covariantly geometrical
field which is necessarily invariant under the holonomy group (and
hence the manifold has a reducible holonomy group). It then
follows that the metric in $M$ is the direct product of the
metrics in $M_M$ and ${\tilde M}_{\tilde M}$ and the manifold has
a product structure $M_N = M_M \times \tilde{M}_{\tilde{M}}$  ($N
= M+ \tilde{M}$), where $M_M$ denotes an M-dimensional Lorentzian
manifold and $\tilde{M}_{\tilde{M}}$ an ${\tilde{M}}$-dimensional
Riemannian manifold \cite{KN}.

Two conformally related manifolds have the same Weyl tensor, and
therefore their algebraic classifications will be equivalent. In
particular, suppose the Lorentzian manifold is an N-dimensional
warped product manifold: $M_N = M_M \times
{_F}\tilde{M}_{\tilde{M}}$,  with metric $g_N = g_M \oplus
{_F}\tilde{g}_{\tilde{M}}$, where the warp factor $F$ is a
function defined on $M_M$. This is conformal to the product
manifold with metric ${\bar g}_N = F\{{\bar g}_M \oplus
\tilde{g}_{\tilde{M}}\}$, and hence the warped manifold has the
same Weyl classification as the conformally related product
manifold. We note that almost all higher dimensional manifolds of
physical interest are either product or warped product manifolds
\cite{examples} (and often are product manifolds obtained by the
simple lifting of a lower dimensional spacetime).

Let us assume that the Weyl tensor is reducible as in (\ref{decomp}), where
${^M}C$ and $\tilde C$  are the $M$- dimensional irreducible Lorentzian and
$N-M$- dimensional Riemannian parts. Then associated with each part would be a
principal type. The principal type of the Lorentzian ${^M}C$ would
be ${\bf G},{\bf I},{\bf II},{\bf III},{\bf N},{\bf O}$. However,
the principal type of the Riemannian
$\tilde C$ is either ${\bf D}$ (type ${\bf II}$, but necessarily all terms of
negative boost weight are also zero) or ${\bf O}$ (the components of
$\tilde C$ are identically zero). We denote the secondary type (ii) of a
reducible Weyl tensor (\ref{decomp}) as $T_{M} \times {\tilde
T}_{\tilde M}$ or simply by $T_{\tilde T}$ if the dimensions $M,
{\tilde M}$ are clear. For example, ${\bf I}_{\bf D}$ denotes a reducible Weyl
tensor in which the irreducible Lorentzian part of the Weyl tensor
is of principal type ${\bf I}$, and the irreducible Riemannian  part of
the Weyl tensor is non-zero. All secondary types are possible, but
we note that while ${\bf G}_{\bf D}$ and ${\bf G}_O$, ${\bf I}_{\bf D}$ and ${\bf I}_O$, 
${\bf II}_{\bf D}$ and ${\bf II}_O$ are of
principal types ${\bf G}, {\bf I}, {\bf II}$, respectively, only types ${\bf III}_{\bf O}, {\bf N}_{\bf O},
{\bf O}_{\bf O}$ are of principal types ${\bf III}, {\bf N}, {\bf O}$ (resp.); secondary types
${\bf III}_{\bf D}, {\bf N}_{\bf D}, {\bf O}_{\bf D}$ are all of principal type ${\bf II}$.

The secondary types of the form (i) and (ii) can of course be
combined with the obvious (albeit somewhat clumsy) notation. Note
that although all of the secondary type (ii)s ${\bf II}_{\bf D}, {\bf III}_{\bf D}, {\bf N}_{\bf D},
{\bf O}_{\bf D}$ are of principal type ${\bf II}$ and the secondary type (i)
${\bf II}_{ii}$ is possible, only secondary type ${\bf II}_{{\bf D},ii}$ is possible
(while secondary types ${\bf III}_{{\bf D},ii}$, ${\bf N}_{{\bf D},ii}$, ${\bf O}_{{\bf D},ii}$ are
not possible).

\noindent
{\bf Full Classification:}
\newline
Alignment type by itself is insufficient for a complete
classification of the Weyl tensor. It is necessary to count
aligned directions, the dimension of the alignment variety, and
the multiplicity of principle directions. We note that unlike in
the four dimensional case, it is possible to have an infinity of
aligned directions. If a WAND is discrete, for consistency with
four dimensional nomenaclature we shall refer to it as a principal
null direction (PND). We can introduce extra normalizations and
obtain further subclasses. The classification of
higher-dimensional Weyl tensors is made more straightforward by
using the notion of reducibility, since we only need classify each
irreducible part (the classification of the irreducible Lorentzian
piece is discussed here and in \cite{MCPP}, and the classification
of the Riemannian piece, which is simpler, is discussed in
\cite{MCPP} and \cite{Karlhede}). However, further complications
in attempting to find canonical forms arise due to gauge fixing
(i.e., certain terms can be chosen to be zero by an appropriate
choice of frame through boosts and spatial rotations).

The {\bf full type} of an irreducible Weyl tensor $C_{abcd}$ is
defined by its principal and secondary types, and includes all of
the information on subcases and multiplicities. We do not classify
these in full detail here (indeed, it may be necessary
to consider different specific dimensions on a case by case basis), 
but rather describe some of the key
algebraically special subtypes below.

First, there are additional conditions for algebraic
specializations: (i) in Type {\bf I} (a)  $C_{010i}=0$, (ii) in
Type {\bf II} (a) $C_{0101}=0$, (b) the traceless Ricci part of
$C_{ijkl} = 0$, (c) the Weyl part of $C_{ijkl} = 0$, (d) $C_{01ij}
= 0$, (iii) Type {\bf III} (a) $C_{011i}=0$. For example, there are consequently
two subcases of Type {\bf III}, namely type {\bf III} (general
type) and type {\bf III} (a) in which $C_{011i}=0$ (see below).
Second, there are further specializations due to multiplicities.
In types {\bf III} and {\bf N} all WANDs are necessarily PND. For
type {\bf III} tensors, the PND of order 2 is unique. There are no
PNDs of order 1, and at most 1 PND of order 0.  For $N=4$ there is
always exactly $1$ PND of order 0. For $N>4$ this PND need not
exist. For type {\bf N} tensors, the order 3 PND is the only PND
of any order.

We can write a canonical form for Weyl type {\bf N}.
From (\ref{eqnweyl}), (\ref{cons}) and (8) we have that for type {\bf N}:
\begin{equation}
C_{abcd} = 4C_{1i1j} \ell_{\{ a} {m^i}_b \ell_c {m^j}_{d\} };
~~C_{1i1}{}^i = 0.
\end{equation}
The general form of the Weyl tensor for type {\bf III} is given by
(\ref{eqnweyl}) subject to (\ref{cons}) and (8). In the subclass {\bf III}$_i$
we have that $C_{1i1j}=0$ and hence
\begin{equation}
C_{abcd} = 8C_{101i} \ell_{\{a} {n}_b \ell
 _c {m^i}_{d\}} + 4C_{1ijk} \ell_{\{a} {m^i}_b {m^j}_c {m^k}_{d
 \}},
\end{equation}
where $C_{011j} = -C_{1ij}{}^i, C_{1(ijk)} = 0$, and not all of
$C_{1ijk}$ are zero (else it reduces to Weyl type ${\bf O}$). In the
subclass {\bf III}(a) we have that $C_{011j} = 0$, so that
\begin{equation}
C_{abcd} = 4C_{1_{ijk}} \ell_{\{a} {m^i}_b {m^j}_c {m^k}_{d
 \}} + 4C_{1i1j} \ell_{\{ a} {m^i}_b \ell_c {m^j}_{d\} },
\end{equation}
where $C_{1ij}{}^i = C_{1(ijk)} = C_{1i1}{}^i = 0$. There is a
further subcase obtained by combining the two cases above
(essentially setting the final term in the last equation to zero).

The complete classification  for $N=4$ is relatively
straightforward, due to the facts that there always exists at least
one aligned direction, that all such aligned directions are
discrete, normalization reduces the possible number of subclasses
(leading to unique subcases) and since reducibility is not an
issue because a Weyl tensor defined over a vector space of
dimension 3 or less must necessarily vanish. In \cite{MCPP} it was
shown that the present classification reduces to the classical
Petrov classification in four dimensions.

It is fortunate that in most applications \cite{wesson,soliton,nrbh,examples,brane} 
the Weyl
classification is relatively simple and the details of the more complete
classification are not necessary. In the future it would be useful
to be able to find a more practical way of determining the
Weyl type, such as for example employing certain scalar
higher dimensional invariants. We also note that it may be more practical in some
situations to classify the Riemann tensor (see \cite{riem} for comments
on the classification of the Riemann tensor in four dimensions)
since it is not
considerably more difficult to classify the Riemann tensor rather
than the Weyl tensor in higher dimensions \cite{MCPP}.

There are many applications of this classification scheme.
Recently we have investigated $N$-dimensional Lorentzian
spacetimes in which all of the~scalar invariants constructed from
the~Riemann tensor and its covariant derivatives are zero (thereby
generalizing the theorem of \cite{cppm} to higher dimensions).
These spacetimes are referred to as vanishing scalar invariant
(VSI) spacetimes, and they can be regarded as higher-dimensional
generalizations of $N$-dimensional pp-wave spacetimes (which are
of interest in the context of string theory in curved
backgrounds). In \cite{cmpppz} we proved that the higher dimensional VSI spacetimes
are necessarily of Weyl type {\bf III}, {\bf N} or {\bf O} (and
we presented a canonical form for the Riemann and Weyl tensors in
a preferred null tetrad in arbitrary dimensions). For
algebraically special vacuum type ${\bf III}$ and ${\bf N}$
spacetimes in arbitrary dimensions we have proven that the null
congruence is geodesic and shear-free \cite{PPCM}. Further progress towards a
generalization of the Goldberg-Sachs theorem is possible in higher
dimensions. In \cite{PPCM}  we also showed that the Weyl tensor in
vacuum type {\bf N} spacetimes with non-vanishing expansion or
twist is always reducible, with a nontrivial four-dimensional
Lorentz part and a vanishing $N-4$ dimensional Euclidean part.

In future work we shall present the details of the full algebraic
classification of higher dimensional Weyl tensors. In particular,
we shall study the five dimensional case in detail (and relate
this classification to previous work \cite{deSmet}), and study type {\bf D}
spacetimes and look for exact solutions that are generalizations
of particular type {\bf D} solutions in four dimensions of special interest
(these include generalizations of the Schwarzschild
solution in higher dimensions, which we note are warped product
manifolds). For example, the higher dimensional spherically
symmetric vacuum non-rotating black hole solutions \cite{nrbh},
which are generalizations of the exact
Schwarzschild solution in four dimensions, are of type {\bf D},
while the five dimensional 
Sorkin-Gross-Perry-Davidson-Owen
soliton solutions \cite{soliton} are all
of (algebraically special) type {\bf I}, except for the special case of a generalized 
Schwarzschild black hole solution which is again of type {\bf D}.

{\em Acknowledgements}. We would like to thank Nicos Pelavas for helpful comments.
AP and VP would like to thank Dalhousie University for its
hospitality while this work was carried out. AC and RM were supported, in
part, by a research grant from NSERC. VP was supported by GACR-202/03/P017.

\end{document}